\documentclass[12pt,aps,prb,preprint]{revtex4}   

\usepackage{amsmath}    
\usepackage{graphicx}   

\begin{document}

\title{Laboratory Exercises Using the Haystack VSRT Interferometer To Teach the Basics of Aperture Synthesis}
\author{J.\ M.\ Marr, A.\ Pere, K.\ Durkota}
 \affiliation{Union College, Schenectady, NY}
 \email{marrj@union.edu}
\author{A.\ E.\ E.\ Rogers, V.\ Fish}
 \affiliation{Haystack Observatory, MIT, MA}
\author{M.\ B.\ Arndt}
 \affiliation{Bridgewater State College, Bridgewater, MA}

\begin{center}
\hspace{1.8in}Submitted to the American Journal of Physics (15-Sep-2011)
\end{center}


\begin{abstract}

    We have developed a set of college level, table-top labs that can be 
performed with an interferometer using satellite TV electronics and 
compact fluorescent lamps as microwave signal sources. This 
interferometer, which was originally developed at the MIT Haystack 
Observatory as a Very Small Radio Telescope (VSRT) to observe the Sun, 
provides students with hands-on experience in the fundamentals of radio 
interferometry. These labs are easily performed and convey an intuitive 
sense of how combining the signals from an array of antennas reveals 
information about the structure of a radio source.

    We have also developed a package of java programs, called 
\textquotedblleft{}VSRTI\_Plotter\textquotedblright{}, 
which is available as a free-download, to facilitate the data 
processing and analysis of these labs.

\end{abstract}


\maketitle

\section{Introduction}

The radio astronomical technique of aperture synthesis has been used for extensive astronomical studies for decades.  Yet, because of the complexity of the math involved, this important technique is almost always excluded from an undergraduate curriculum in physics and astronomy, leaving the interested students mystified about how instruments like the Very Large Array in New Mexico\cite{VLA} actually work.

We present here a discussion of a set of labs for the undergraduate level which provide first-hand experience with the basics of making images with aperture synthesis.
Although a thorough mathematical explanation of the process would be helpful for the advanced students, these labs are designed to be stand-alone so that they can be completed without the mathematical lectures and still impart a general conceptual understanding to the lower level students. 

\section{Some Basics of Aperture Synthesis}

A number of antennas arranged in an array with assorted baselines receive the radiation from a given source simultaneously, and the output signals from each pair of antennas are cross-correlated.  Correcting for a few systematics, the cross-correlations
lead to what is known as the \textquotedblleft{}Visibility Function,\textquotedblright{} which is a complex-valued function of the baseline vector.
An image of the source is then obtained via the Fourier transform of the Visibility function.

The Visibility for any given baseline vector between antennas contains an amplitude and a phase and so is simply represented by
\begin{equation}
\label{eq:Vis}
V(\vec b ) = Ae^{i\phi},
\end{equation}
where $\vec b$ is the projection of the baseline onto the sky plane.
For the purposes of this paper, we can simplify this discussion and consider just the two-dimensional situation in which the source structure and the antennas are located in a single plane, in which case the baselines and the source structure are each one-dimensional.  

For a number of point sources, the visibilties due to all the individual sources simply add, so that
the total Visibility for baseline b is
\begin{equation}
V_T(b) = Ae^{i\phi} = \sum_k A_k e^{i\phi_k}
\end{equation}
The magnitude of any particular Visibility, given as the square root of the sum of the squares of the total real part and the total imaginary part, then, is given by\cite{HowVSRTworks}
\begin{equation}
\label{eq:amp}
A = \sqrt{\sum_k A_k^2 + \sum_{l>k} 2 A_k A_l \cos(\phi_k - \phi_l)}.
\end{equation}

Since the Visibility function is the Fourier transform of the image,
the shape of the Visibility function contains information about the structural aspects of the source.  For example, if observing a pair of equally bright point sources, the 
Visibility function has nulls whose spacing in projected baseline is 
inversely proportional to the angular distance between the point sources. 
More specifically, the nulls are at
\begin{equation}
b \Delta\theta = N ({\lambda \over 2})
\end{equation}
where N is an odd integer, b is the projected baseline and $\Delta\theta$ is the angular separation of the point sources.

And, if observing a single but resolved source, the Visibility function will be
a decreasing function with baseline length and the rate of decrease will be inversely proportional to the angular size of the source.  For a source with a Gaussian brightness profile, for example, the Visibility function will also decrease with a Gaussian profile, and the half-maximum width of the Visibility function will be inversely proportional to the half-maximum width of the source brightness distribution.

In the labs we discuss here, the students obtain a first hand exposure to the nature of the Visibility function and how it relates to the source structure.  Without any discussion of Fourier transforms, or even complex numbers, students gain an intuitive understanding of how an array of radio antennas produces images of radio sources.

\section{The VSRT Interferometer}

The labs we discuss here use the Haystack VSRT (Very Small Radio Telescope) Interferometer.
This instrument costs less than \$500, is easy to
assemble, is stable and reliable, is easy to operate, and is easily manipulated in
the lab room.  Information about purchasing and assembling the VRST
interferometer can be found at http://www.haystack.mit.edu/edu/undergrad/VSRT/index.html.  

The VSRTs come with satellite TV dishes, which are needed for observations of the Sun, as described by Doherty, Fish, and Needles (2011)\cite{DFN}.  But, for the labs in the classroom the dishes are not needed and so in all the labs we discuss
here the feeds act as the antennas.  In the following, then, the words \textquotedblleft{}feeds\textquotedblright{} and \textquotedblleft{}antennas\textquotedblright{} are interchangeable.

The radio sources to be used in the lab room with the VSRT interferometer
are, simply, compact flourescent light bulbs (CFLs). In addition to
visible light, these light bulbs emit microwaves, which can be detected
by the VSRT feeds. This radio emission is due to bremsstrahlung radiation
emitted by hot, free electrons in the plasma, produced by the gas discharge,
when they collide with the glass walls.

\subsection{How the VSRT Interferometer Works}

The VSRT interferometer differs from modern interferometers
used by radio astronomers today in two significant ways. First,
the signals are added instead of cross correlated (i.e., it is an
additive interferometer) and, secondly, the two feeds involve different
mix-down (i.e. Local Oscillator, or \textquoteleft{}LO\textquoteright{}) frequencies. 
Therefore, when the signals are combined a beat
signal (with frequency about 500 kHz) results.
The end result, however, is that the response of the VSRT
interferometer mimics, in many ways, a standard cross-correlation interferometer,
as we show below.

First consider the radiation from a single source entering the two feeds
and the detected power that exits the square-law detector.  We'll
denote the baseline distance between the feeds as \textquoteleft{}b,\textquoteright{} and, 
assuming that the distance of the source is much
larger than b, we assign the position of the source by its direction
angle, $\theta$, relative to the mid-plane between
the feeds.
The electric field entering each feed is 
\begin{equation}
E_a = E_0 ~\cos(2\pi\nu t) ~and~ E_b = E_0 ~\cos(2\pi\nu t - \phi),
\end{equation}
where the phase difference, $\phi$, is due to the extra path length to the second antenna and is given by
\begin{equation}
\phi = 2\pi {b\sin\theta \over \lambda}.
\end{equation}
The signals are then mixed down with frequencies
$\nu_a$ and $\nu_b$, added, and squared with only low-frequencies passing through.  The end result
of all these steps is\cite{HowVSRTworks}
\begin{equation}
V_T^2=V_0^2\cos[2\pi(\nu_b-\nu_a)t+\phi]. 
\end{equation}
This, simply, is the beat signal.
For simplicity of notation, we will write the amplitude of the beat signal as \textquoteleft{}S\textquoteright{}, and so the last equation becomes
\begin{equation}
\label{eq:output}
S=S_0\cos[2\pi(\nu_b-\nu_a)t+\phi]. 
\end{equation}

Now consider the total power in the beat signal when there are N sources.
In general, the sources are incoherent and so we must 
add the powers.  The total power in the beat signals of N sources, then, is given by  
\begin{equation}
\label{eq:Nsources}
S_T=\sum_k S_k \cos[2\pi(\nu_b-\nu_a)t+\phi_k],
\end{equation}
where $S_k$ is the power in the beat signal from the kth source and $\phi_k$
is the phase delay to the second feed from the kth source.  This is mathematically identical to\cite{HowVSRTworks}
\begin{equation}
\label{eq:VSRTamp}
S_T = \left( \sqrt{ \sum S_k^2 + \sum_{l>k} 2 S_k S_l \cos(\phi_k - \phi_l)} \right) \,\, \cos(2\pi(\nu_b-\nu_a)t).
\end{equation}
Note that this, again, is the beat signal, but with an amplitude that is
modified by the factor with the radical.  Note also that the amplitude of the beat signal
in Eq.~\ref{eq:VSRTamp} is identical to 
Eq.~\ref{eq:amp}.  We see, therefore, that the power of the beat signal with the VSRT interferometer is identical to 
the amplitude of the complex visibility for that baseline for any arrangement of sources.

\section{The Undergraduate Labs}

Below we discuss a sequence of four labs designed to impart an intuitive sense of how data from
an array of antennas can produce images of radio sources.  Depending on the goals and conditions of the class the instructor could choose to skip
the last lab and still successfully convey the basics of the relation between the source structure and the data between pairs of antennas. 

\begin{figure}[h]
\scalebox{0.12}{\includegraphics{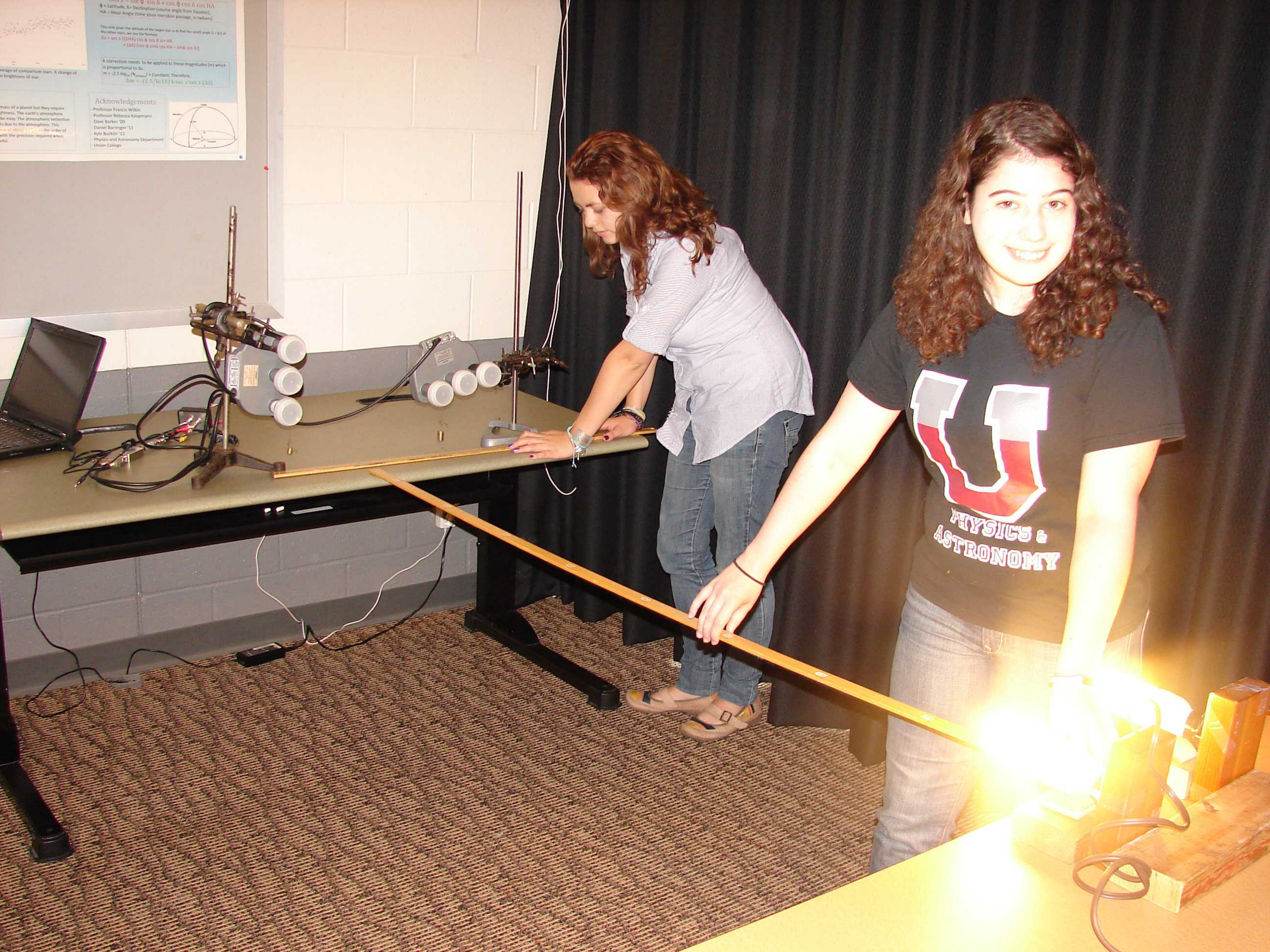}}
\caption{\label{fig:setup}Undergraduate students from the Union College radio astronomy independent study group work with the VSRT interferometer.  Two feeds taken from TV satellite dishes,
are aimed at two CFL's, which act as radio sources.  The feeds shown are \textquotedblleft{}triple\textquotedblright{} DirecTV feeds for receiving 3 satellites at 101, 110, and 119$^o$ West in geostationary orbit.  The lowest feed of each triple as shown is the active feed.  The nominal local oscillator frequency is 11.25 GHz for reception fo the 11.7 1 12.2 GHz band.  The feed polarization is left circular (LCP) which becomes right circular (RCP) upon reflection from the dish when used for TV.
The distance between the feeds is determined by reading their positions relative to the meter stick below them.  The output spectrum from the interferometer is displayed on the laptop screen, and the data files are recorded.  The CFL's, which are easily movable, are located two-meters from feeds, which can be moved to assorted separation distances.}
\end{figure}

For the sake of simplicity the lab set-ups are two-dimensional.  Maps of the sources, then, are merely plots of intensity vs.\ position in the horizontal direction.
In Figure~\ref{fig:setup} an example of the set-up is shown.   

Complete instructions for all of these labs is also available at https://www1.union.edu/marrj/radioastro/labfiles.html.

\subsection{Analysis Software}

To facilitate the analysis of the data for these labs we have produced a 
package of java programs into which the output data from the VSRTI inteferometer
control program are easily fed.  These programs also have links to the lab instructions
and can produce overlays of theoretical models with
adjustable parameters.  

We have called this package \textquotedblleft{}VSRTI\_Plotter\textquotedblright{}.  A
free and downloadable zip file of the VSRTI\_Plotter package is located at 
https://www1.union.edu/marrj/radioastro/labfiles.html.

\subsection{Lab \#1: The VSRT Primary Beam}

In the first lab, the students use the VSRT interferometer to observe a single CFL at varying angular positions
and discover that the detected power depends on the position of the CFL.  They learn about the \textquotedblleft{}primary beam\textquotedblright{}
of the individual antennas and that when sources are not at the center
of the primary beam, the detected power is decreased by
the primary beam factor.  This is an exercise that's relevant to single-dish radio astronomy
as well, since the beam size of the antenna determines the resolution of the telescope
and the step size that a single-dish telescope must move to map a source.  
Additionally, the size of the primary beam places an effective upper limit on the
maximum field of view in an aperture synthesis observation with a single pointing.
Knowledge of the primary beam pattern is also important to the analysis in
the fourth lab.

The VSRTI\_Plotter program enables the students to overlay a theoretical beam plot and obtain a fitted value of the diameter of each antenna.

\begin{figure}[h]
\scalebox{0.6}{\includegraphics{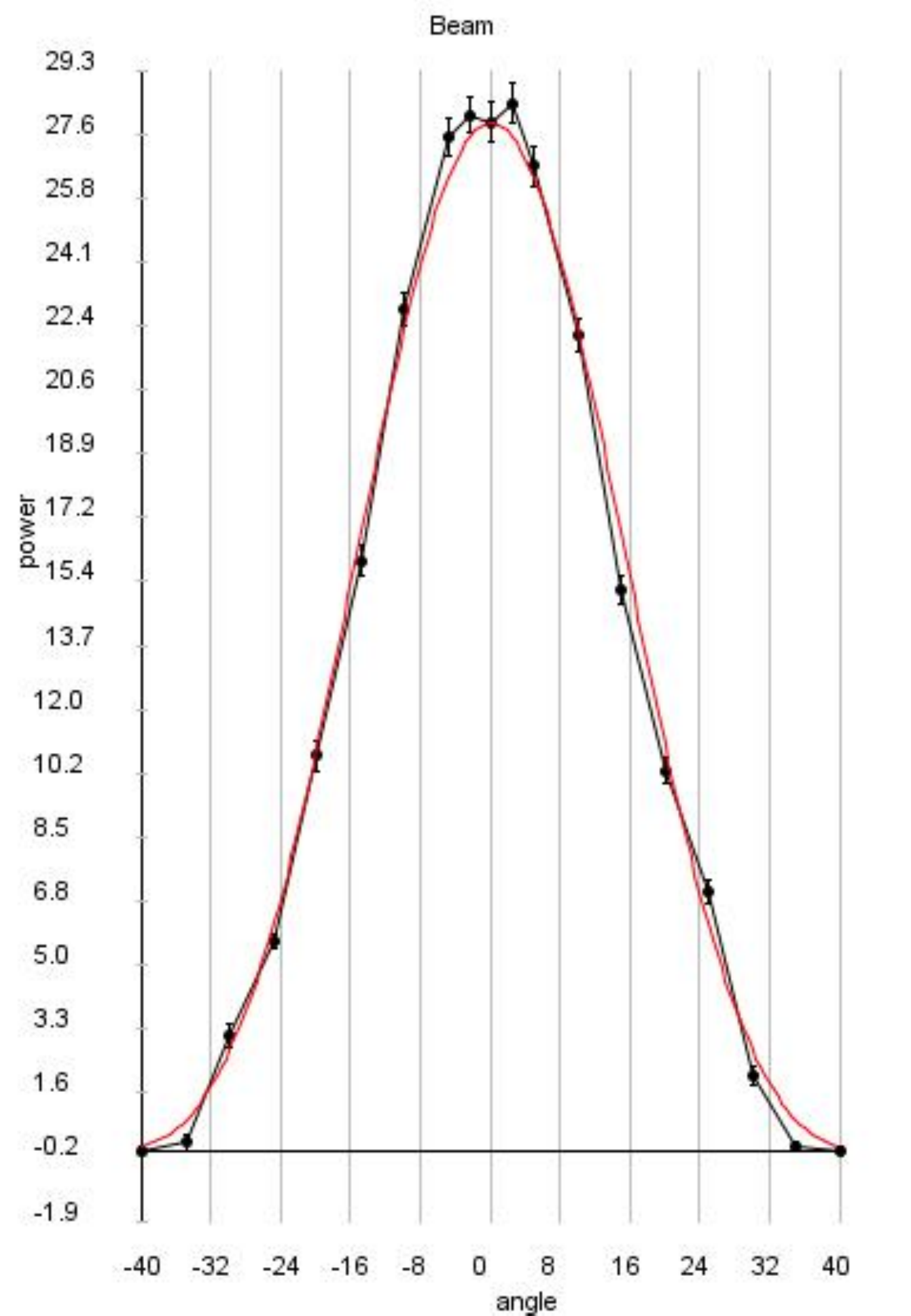}}
\caption{\label{fig:beamplot}A plot of detected power vs.\ angular position of a CFL, revealing the VSRT's primary beam. The model overlay yields a measure of the effective diameter of the antenna feed of approximately 3.8 cm.}
\end{figure}

In Figure~\ref{fig:beamplot} we show the beam pattern plot displayed in VSRTI\_Plotter with data 
obtained by undergraduate students in the radio astronomy class at Union College. 

\subsection {Lab \#2: The Visibility Function of a Single Resolved Source:}

\begin{figure}[h]
\scalebox{0.6}{\includegraphics{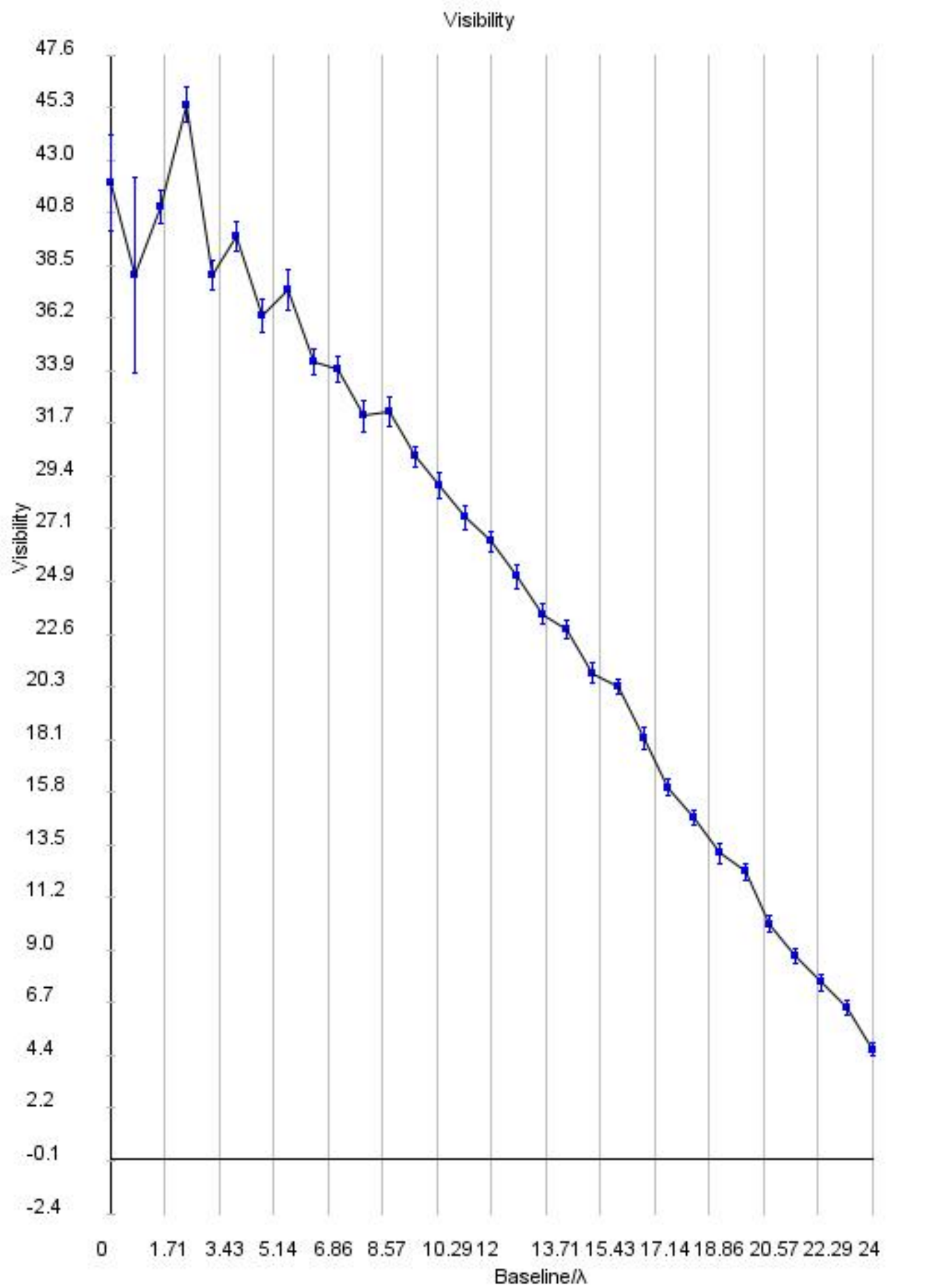}}
\caption{\label{fig:singlesource}The detected power when observing a single CFL 
vs.\ the separation distance between the feeds.  The plot shows
that with increasing baseline lengths more of the flux of the source
gets resolved out.  The smooth curve represents the model in which the brightness
and angular width of the source is adjusted to fit.
The same measurements can be made with two CFL's positioned side-by-side to model
a single source twice as wide.  Students then discover that the detected power decreases
twice as fast.}
\end{figure}

Using a single CFL, the students are instructed to vary the separation distance between
the feeds and plot the measured power vs.\ baseline distance and so
are introduced to the \textquotedblleft{}Visilibity function.\textquotedblright{}  
Figure~\ref{fig:singlesource} displays data obtained by undergraduate students.
The students see that the detected signal decreases with baseline distance.
They then place two CFL's side-by-side, to simulate a source twice as large and discover that the
decrease in the Visibility function is faster when the source is wider.  
Without delving into the mathematics
of the Visibiliity function, the students gain the intuition that a plot of interferometer
response vs.\ baseline distance contains information about the angular size of the 
source.  

With the overlay model option in VSRTI\_Plotter, the students can fit the observed data and find
a best-fit measure of the angular size of the source and see that it agrees with that expected.

\subsection{Lab \#3: Visibility Function for a Pair of Sources}

\begin{figure}[h]
\scalebox{0.6}{\includegraphics{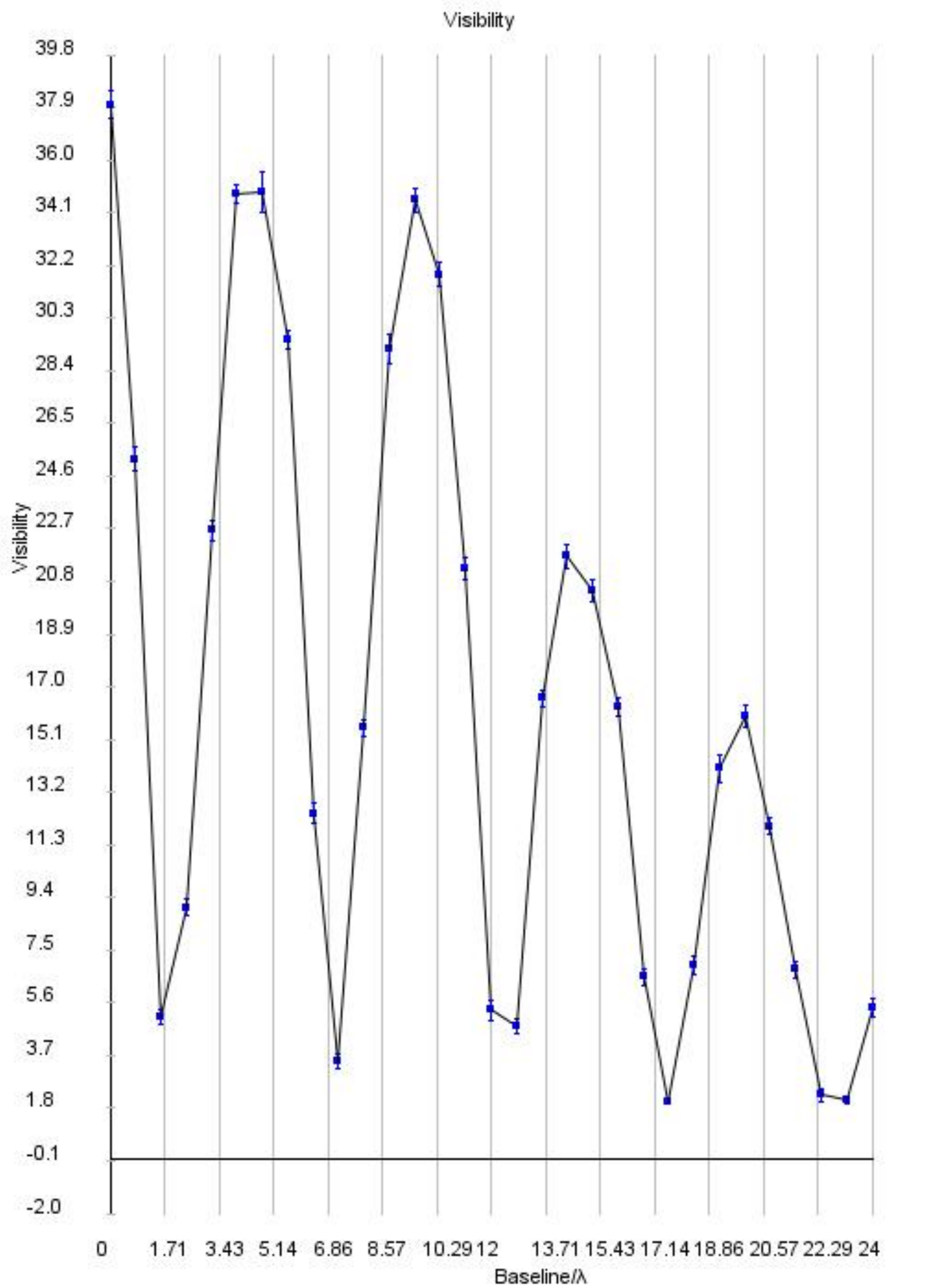}}
\caption{\label{fig:doublesource} The detected power with a pair of CFLs vs.\
baseline length.  The detected signal has
an oscillating dependence on baseline length.  The overall decrease
with baseline is due to the single-resolved source pattern of Lab \#2.
The experiment can be repeated for different CFL separations
to demonstrate that the periodicity of the oscillating function is
inversely related to the separation angle of the sources.}
\end{figure}

Here, the students observe two CFLs separated by a given distance 
and again observe with the feeds at varying separations.  The students
discover that
when observing a double source the visibility function oscillates.  
Data for this lab obtained by students are displayed
in Figure~\ref{fig:doublesource}.  By
changing the separation distance of the CFLs and re-observing with the same
set of baselines they discover that the oscillation length of
the Visibility function is 
inversely related to the separation distance of the two sources.

Again, with VSRTI\_Plotter, the students can fit a model to the data to obtain measures 
of the angular separation of the sources and compare with their set-ups in the lab.  

\subsection {Lab \#4: Examination of Interferometer \textquotedblleft{}Fringe Pattern\textquotedblright{}}

In this final lab, the students discover the 
\textquotedblleft{}fringe pattern\textquotedblright{} of
an interferometer.  Although for some classes the principle contained in this lab 
may seem too esoteric, for students with strong backgrounds likely
to continue on in astronomy, or for students starting research
projects in radio astronomy, this lab exercise provides a visual 
demonstration of a fundamentally important concept that newcomers to
aperture synthesis commonly find confusing.

When an interferometer is used to observe
a celestial point source the direction of the 
source changes continuously as the Earth rotates causing
the resulting detected power vs.\ time to oscillate quasi-sinusoidally.\cite{TMS,BGS,WRH}
The \textquotedblleft{}fringe frequency\textquotedblright{} depends on the source position and
is greatest when the source is near transit for an East-West baseline.  One can also speak of
the \textquotedblleft{}fringe pattern,\textquotedblright{} in which the detected signal
is mapped as a function of the source's position in the sky.
The fringe pattern is important to interferometery in three ways. 
First, fitting this function to the detected power vs.\ time yields an
accurate measure of a source's position in the sky.
Secondly, this function is used to calibrate interferometer baselines by observing
a bright point source whose position is well known and fits the
fringe function. 
And, thirdly, this function is used
to determine when a faint source is being detected. 
If the power has a time dependence that fits the fringe function
then the observer knows that it is due to a celestial source.

To reproduce the fringe pattern with the VSRT interferometer, since the VSRT interferometer
cannot measure the fringe phases directly, one must place and fix one CFL at $\theta=0$ as
a reference source, while the fringes due to a second CFL is detected.  Additionally, 
since the source doesn't naturally move relative to the baseline, the students must manually
move the second CFL to various positions.  While leaving the feeds stationary at a given
baseline and one CFL at the center position, the students record data with the second CFL
located at a range of angles at a constant distance to the mid-point between the feeds.  The detected power, then, is given by Eq.~\ref{eq:VSRTamp} with just two sources, $\theta_1 = 0$ and $\theta_2 = \theta$. 
\begin{equation}
\label{eq:fringepattern}
S = \sqrt{S_1^2 + S_2^2 + 2 S_1 S_2 \cos(\phi)}
\end{equation}

\begin{figure}[h]
\scalebox{0.6}{\includegraphics{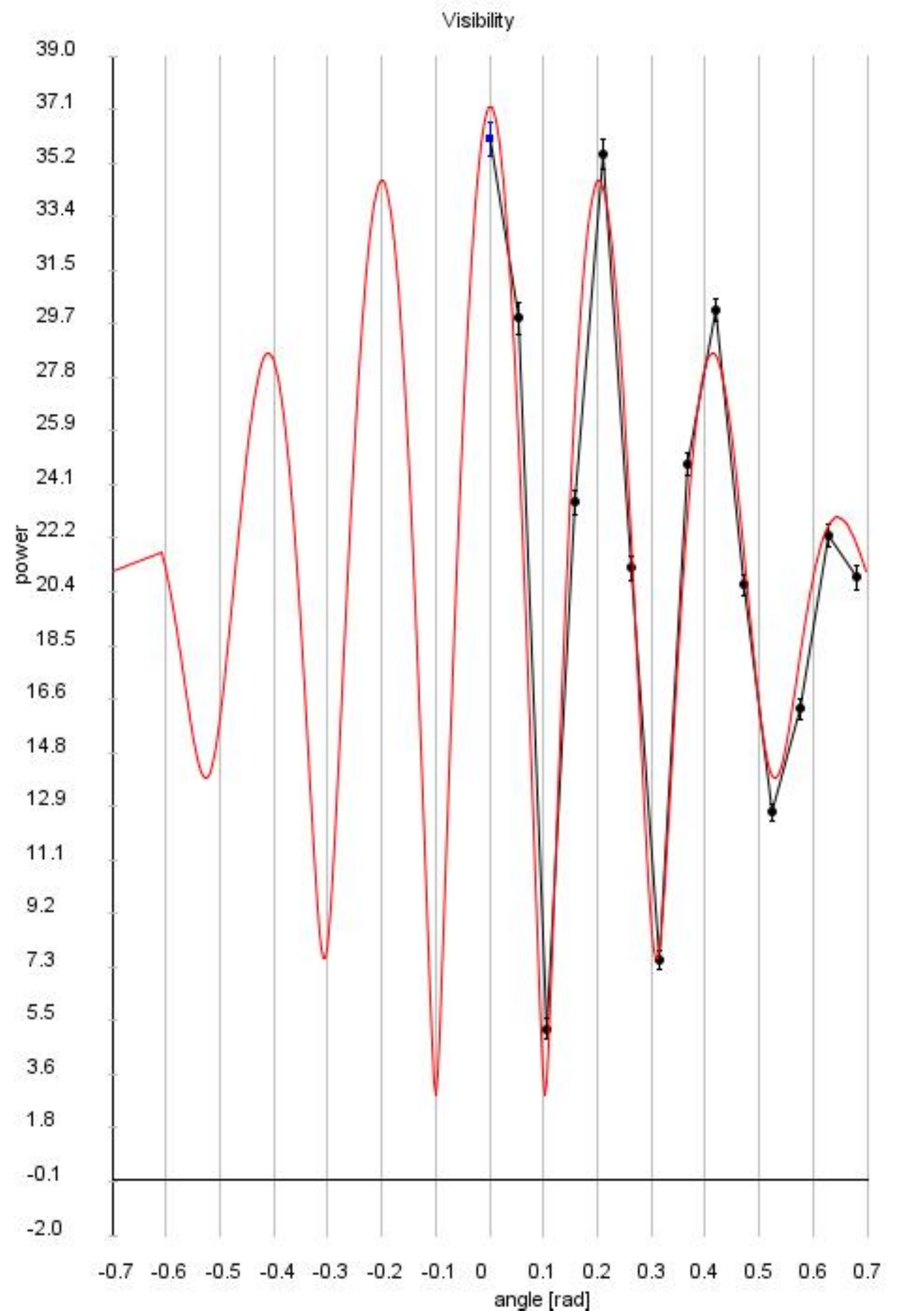}}
\caption{\label{fig:fringepattern}A plot of the detected power vs.\ angular position of a CFL, while another CFL stays fixed at $\theta=0$.  The resultant plot reveals the \textquotedblleft{}fringe pattern,\textquotedblright{}, although with an outer  
envelope due to the primary beam pattern.  The model overlay, obtained for a baseline of 
12.5 cm, more clearly shows how the fringe rate depends on the source position.  Note that the distance between peaks when the source is near $\theta=0$ is 0.2 radians, while at larger angles the peaks occur at 0.4 and 0.65 radians.}
\end{figure}

A data set for this lab obtained by undergraduate students is shown in Figure~\ref{fig:fringepattern}. 
The amplitude of the oscillations decrease toward larger angles because the CFL moves out of the center of the primary beam.

\section{Concluding Remarks}

A group of undergraduate students undertaking an independent study on radio astronomy at Union College in Fall 2010 completed the set of labs discussed above.  In an independent study these students would not have succeeded in digging through the many pages of high level math needed to understand
the basics of aperture synthesis and so these students would have completed their study considering only single-dish observing methods.  Instead, these students
ended the term with first-hand experience that demonstrated that obtaining data between 
pairs of antennas at many different separation distances leads to a function from which one could infer
the size of a single source or the separation of a pair of sources.  With a little further discussion,
and possibly demonstrations using images from actual aperture synthesis data,
an understanding of the concept that the structural details of even a complex source can be extracted from such data 
straightforwardly using established algorithms is obtained.

\begin{acknowledgments}

We are grateful for the significant contributions by Preethi Pratap and Madeleine Needles in
the development of the VSRT, for the helpful assistance of Francis Wilkin in the development of the labs, and for the careful lab work of the Union College radio astronomy
students, namely Daniel Barringer, Halley Darling, Ana Mikler, and Katelyn O'Brien.

The development of the VSRT was funded by the NSF through a CCLI grant to MIT Haystack Observatory.  
The development of the VSRTI\_Plotter java package was funded by the National Science Foundation, IIS CPATH Award \#0722203

\end{acknowledgments}


\begin{thebibliography}{5}

\bibitem{VLA}The Very Large Array is operated by the National Radio Astronomy
Observatory, which is a facility of the National Science Foundation
operated under cooperative agreement by Associated Universities, Inc.

\bibitem{HowVSRTworks} Derivations are shown at https://www1.union.edu/marrj/radioastro/HowtheVSRTworks.pdf

\bibitem{DFN}Doherty, M., Fish, V.\ L.\ and Needles, M.,\textquotedblleft{}Revealing the Hidden Wave: Using the Very Small Radio Telescope to Teach High-School Physics,\textquotedblright{} Physics Teacher, in press (2011)

\bibitem{BGS}Burke, B.\ F.\ and Graham-Smith, F., \textsl{An Introduction to Radio Astronomy}
(Cambridge University Press, Cambridge, UK, 2010), 3rd ed.

\bibitem{TMS}Thompson, A.\ R., Moran, J.\ M., and Swenson, Jr., G.\ W., \textsl{Interferometry
and Synthesis in Radio Astronomy} (John Wiley \& Sons, Inc., New York, NY, 2001), 2nd ed.

\bibitem{WRH}Wilson, T.\ L., Rohlfs, K., and H\"{u}ttemeister, S., 2009 \textsl{Tools of Radio
Astronomy} (Springer-Verlag, Berlin-Heidelberg, 2009), 5th ed.

\end{thebibliography}
\end{document}